**Improving Resolution and Resolvability of Single Particle CryoEM Structures using Gaussian Mixture Models**

Muyuan Chen[1]*, Michael F. Schmid[1], Wah Chiu[1,2]

[1] Division of CryoEM and Bioimaging, SSRL, SLAC National Accelerator Laboratory, Stanford University, Menlo Park, CA 94025, USA

[2] Department of Bioengineering, and of Microbiology and Immunology, Stanford University, Stanford, CA 94305, USA

* Correspondence to: muyuanc@stanford.edu

**Abstract**

Cryogenic electron microscopy is widely used in structural biology, but its resolution is often limited by the dynamics of the macromolecule. Here, we developed a refinement protocol based on Gaussian mixture models that integrates particle orientation and conformation estimation, and improves the alignment for flexible domains of protein structures. We demonstrated this protocol on multiple datasets, resulting in improved resolution and resolvability, locally and globally, by visual and quantitative measures.

**Main**

Single particle cryogenic electron microscopy (CryoEM) leverages information from multiple independent observations of the same protein at different orientations to reconstruct its structure in 3D. This method can now determine rigid protein structures at atomic resolution[1–4], and the number of near-atomic resolution CryoEM structures deposited in the Protein Data Bank continues to rise. In order to reconstruct the 3D structure, the orientation of 2D particle images are determined computationally, typically by iteratively searching for the optimal pose (rotation-translation parameters) that maximizes the similarity between each particle image and the projection of the 3D density map[5,6].

To estimate the pose of particles, algorithms assume that their 2D images represent observations of the same 3D structure at different orientations. However, this assumption becomes invalid when the macromolecule has compositional or conformational heterogeneity, leading to limitations in the resolution achieved by CryoEM. To overcome this issue, various computational methods have been developed. Focused and multibody refinement addresses heterogeneity by refining particle orientation for individual domains after density subtraction, but the alignment can be affected by artifacts from masking and subtraction[7]. Machine learning-based manifold embedding methods have been proposed to address protein flexibility by learning the conformational changes from particle images[8,9]. However, it remains challenging to obtain high-resolution reconstructions for the flexible regions, because the signal to noise ratio (SNR) is often limited by the number of particles at the exact same conformation along a continuous movement trajectory. While the learned conformational changes can sometimes be used to generate deformed structures with improved features[10,11], they are different from direct reconstruction from particles, which are free from deformation artifacts and may contain features not present in the consensus map.

Recently, we have developed a protocol that uses Gaussian mixture models (GMMs) for structure representation, and extract movement trajectories of proteins directly from the particles using deep neural networks (DNN) techniques[12]. Instead of using voxel maps, GMMs represent protein structures as the sum of Gaussian functions, which is a natural way to depict molecular models. To further improve the resolution and resolvability of the map and model, we expanded the GMM-based refinement workflow, which now optimizes the poses of particles and their conformations (**Fig.S1**). Using this technique, the knowledge of conformational heterogeneity can be converted to the poses of particles, to better resolve domains undergoing large-scale continuous movement.

For GMM-based orientation refinement, the protocol starts with a homogeneous refinement and uses existing pose estimations provided by voxel-based methods using software such as Relion[5], EMAN2[6], and CryoSparc[13]. To prevent overfitting, the "gold-standard" even/odd particle split is kept throughout the entire process, and two GMMs are independently built from the two half sets. In each iteration, the pose of particles are locally refined using a gradient-based optimizer[14], and 3D reconstructions are computed from the particle images with new orientations using direct Fourier inversion onto a voxel grid. The resolution of reconstructions is measured by "gold-standard" Fourier shell correlation (FSC) between the two voxel-based half maps. The usage of GMM brings a number of advantages. First, compared to the voxel maps, the GMMs require only ~1% of variables to represent the same structure[15], which reduces the computational cost and increases the observation/variable ratio. Second, as we only project the Gaussian coordinates to generate Fourier images, no interpolation is needed throughout the alignment. Third, since central

Gaussian coordinates are located on protein densities, the GMM essentially provides a tight mask around the molecule without creating artifacts in real or Fourier space. Finally, because the GMM is a continuous function in 3D, the optimization can take advantage of accurate gradients calculated using deep learning packages[16]. All these features are particularly helpful for retrieving high resolution signal with low SNR, without the artifacts from interpolation and masking, which can have a strong impact on alignment.

Next, we implement GMM-based focused refinement. With GMMs, we can adjust not only the amplitude, but also the width of Gaussian functions within a region. Since Gaussian width represents the local resolvability, by increasing the width of the GMM references outside the region, the alignment will focus on high-resolution features within the region, while still being constrained by the low-resolution information from the rest of the structure. Furthermore, we can use a patch-by-patch refinement strategy to improve the resolvability of the entire macromolecule. That is, we segment the full GMM into multiple patches, focus refine each individual one, then merge the results together into a final composite map.

Finally, we integrate the pose estimation with the previously developed heterogeneity analysis[12]. The process starts from the DNN-based heterogeneity analysis using the orientation assignment from the GMM-based consensus refinement. After the heterogeneity analysis, the trained decoder outputs one GMM for each point from the latent space, corresponding to the conformation of that particle. While the overall movement within a protein can be complex, we assume that within a small enough local region, the structure moves as a rigid body. Therefore, we can change the frame of reference to focus the alignment on the target region. For each particle, we optimize the pose to minimizes the difference within the region between the GMM of the conformational state of that particle, and the neutral state GMM (i.e., GMM derived from the consensus refinement) at the initial orientation. Reconstruction of all particles at the new orientation will have better-resolved features at the target region, and smeared-out densities in the rest of the structure. A few rounds of focused refinement are then performed starting from this Euler angle assignment to further improve the resolvability of that region.

To demonstrate the performance of the method, we tested it on three datasets. To make a fair comparison, we selected public datasets with assigned particle orientations provided by different authors. In all examples, the same particles, masking, and structure sharpening were used, to ensure the only difference between the results comes from the orientation assignment.

Our first test was of a relatively rigid GPCR dataset from EMPIAR-10786, which was initially determined at 3.3Å "gold-standard" resolution using multiple software with global and focused refinement[17]. The overall structure was improved to 2.5Å after GMM-based patch-by-patch refinement. As a measure of resolvability, average Q-score[18] increased from 0.60 to 0.67. The GMM-based refinement improves real space features across the entire density map. At the nanobody, β-strands are better separated and previously unmodeled broken density becomes connected. Along the ECL at the opposite side of the protein, we also observed better backbone connectivity and side chain density (**Fig.1A-D, Fig.S2, Supplementary movie 1**).

Next, we demonstrated focused refinement on a SARS-COV2 spike dataset (EMPIAR-10492)[19]. In this dataset, the central parts of the spike are well resolved, whereas the upper parts, including the RBD and the NTD are more flexible. Here, we improved "gold-standard" resolution from 3.8Å to 3.1Å, and the average Q-score from 0.47 to 0.53. The refinement also greatly improved features at the flexible domains. Strands in β-sheets become clearly separable, and side chains start to show up in the refined structure (**Fig.1E-H, Fig.S3, Supplementary movie 2**).

In the third example, we show the integration with the DNN-based heterogeneity analysis using an ABC transporter dataset (EMPIAR-10374)[20]. This structure contains rigid TMDs, flexible NBDs, and Fabs that are undergoing large-scale movement and are blurred out in the consensus map. The global "gold-standard" resolution is 3.2Å, but the Fab is only determined at 5-10Å resolution locally. The global GMM-based refinement increased resolution to 2.8Å. Focused refinement improved the resolvability of the NBD, but its improvement on the Fab was suboptimal. So, we ran the DNN-based heterogeneity analysis focusing on one of the Fabs, which showed a continuous tilting movement of as much as 14Å. Then, the conformations of particles were converted to the change of pose at that region, and a few rounds of focused refinement were performed afterwards. The refinement yielded clear high-resolution features at the Fab, including separable β-strands with side chains. Combining with patch-by-patch refinement results for the rest of the protein, the overall resolution reached 2.5Å (**Fig.2, Fig.S4, Supplementary movie 3, 4**).

In summary, by refining particle poses using GMMs instead of voxel maps as references, we obtain reconstructions with higher "gold-standard" resolution and better resolved real-space features. The use of GMMs also makes it easier to focus the alignment on local regions, and by segmenting the GMM into patches and focusing on every patch individually, we can further improve the resolution and resolvability of features throughout the structure. Combined with the DNN-based heterogeneity analysis, information of conformational changes can be converted to particle poses of the target domain, leading to greatly improved features at the highly flexible protein domains that are undergoing large-scale continuous motion. In addition to the near-atomic resolution structures from the examples, the GMM-based refinement also shows clear improvement for structures initially determined at medium resolution (see Methods and **Fig.S8**). The method is compatible with most CryoEM software packages, and shows superior performance on a wide range of macromolecules.

# Methods

## Introduction

The basic concepts of GMMs, as well as the architecture and training process of the DNNs, have been described in our previous publications[12,15]. In this section, we briefly explain the basic concepts of the method, but it is recommended to refer to the previous papers for more details.

To represent the structure of proteins, we use a GMM, which is a sum of Gaussian functions in real space, $\bar{x} \in R^3$. Each Gaussian function is represented by five parameters: amplitude $A_j$, width $\sigma_j$, and the 3D center coordinates $\bar{c}_j$.

$$M(\bar{x}) = \sum_{j=1}^{N} A_j e^{-\frac{|\bar{x}-\bar{c}_j|^2}{2\sigma_j^2}}$$

To generate 2D projections of a GMM, we project the coordinates of each Gaussian function using the rotation matrix $R$, and directly generate the 2D image using the Gaussian parameters. Since we only project coordinates, no interpolation is involved in the process. A projection of the GMM in $\bar{t} \in R^2$ is simply:

$$P(\bar{t}) = \sum_{j=1}^{N} A_j e^{-\frac{|\bar{t}-R\bar{c}_j|^2}{2\sigma_j^2}}$$

To compare projections of the GMM and particle images, we use the Fourier ring correlation (FRC) as loss function. The FRC between the Fourier transform of the GMM projection and a CryoEM particle image, is the average of the correlation coefficients over Fourier rings. Since each ring is independently normalized, the FRC is insensitive to filtering of images. As long as the CTF phases have been correctly flipped, CTF amplitude correction can be ignored. During the pose and conformation estimation, the FRC will be maximized for an individual particle when the GMM best agrees with the underlying particle density irrespective of CTF.

The optimization of GMMs is performed using DNNs. To build a GMM that matches a given voxel-based density map, we make projections of the density map, and train the DNNs to adjust the parameters of the GMM so its projections match the projection of the given density map at the same orientations. The DNN used here is a densely connected network with three layers. Details about the DNN architecture and training procedure can be found in the previous paper[12]. While the DNN can be trained from scratch, when initial locations of Gaussian coordinates are provided, it can also be pre-trained so the coordinates of the Gaussian functions in the GMM match the given 3D locations. The full GMM can then be optimized to match the density map using the initial coordinates as a starting point.

For heterogeneity analysis, a pair of DNNs, one encoder and one decoder, is built from the particles. The encoder takes the information of each particle and maps it to a latent space representing the conformational state of the particle. The decoder network then takes the coordinates in the latent space and produces a GMM at the corresponding conformation. While we always use the densely connected network for the encoder, multiple DNN architectures for the decoder have been established[12,15], and users can choose the architecture using command line options. By default, and in examples shown in this paper, we use the hierarchical GMM described previously[15]. The small GMM used in the hierarchical DNN contains 32 Gaussian functions, which is used to drive the movement of the full GMM in the first round of training. The location of Gaussian functions in the small GMM is determined by K-means clustering of the full GMM. Details on the DNN design and training procedure can be found in the previous paper[15].

**Construction of Gaussian model**

The first step in constructing the GMM is to determine the number of Gaussians (N) needed. The recent improvement of the GMM implementation makes it possible to represent the macromolecule with tens of thousands of Gaussian functions within the capability of modern GPUs[15]. So, when the sequence is known and the target resolution is 3Å or higher, it is convenient to simply set N to be the number of non-H atoms in the molecule. In our experience, small changes (10%) of N does not have a clear impact on the result of the refinement. For reconstruction at lower resolution, we also provide tools to automatically estimate the necessary number of Gaussian based on the size of protein and target resolution. In this approach, we build many GMMs of variable N from the same given density map, and determine the optimal N empirically. The goal here is to have the average nearest neighbor distance between Gaussian functions of the GMM match the value of the target resolution. Since resolution can be defined as the minimum distance that two Gaussian functions can be separated, this approach can provide a good estimation of N to generate a GMM that represents the target structure.

With a given number of Gaussian, there are two ways to initialize the location of Gaussian functions in the GMM. First, when an existing molecular model is available, we can take the coordinates from the PDB file, and seed one Gaussian per non-H atom. To avoid overfitting, the coordinates of each Gaussian in the GMM is randomized at the determined "gold-standard" resolution before being used as references for the even/odd subsets. Alternatively, the initial location of Gaussian functions can be determined by clustering voxels in the given density map with values above a given threshold using the K-means algorithm. Empirically, the two initialization methods cause little difference in the results of particle orientation refinement. In our experiments, the two methods yield overlapping FSC curves and virtually identical real space features after the refinement.

However, there are some potential points of concern with both these initialization methods. Seeding the GMM with a molecular model can lead to incomplete connectivity of polypeptide if the given atomic model was derived initially from a poorly resolved density map and thus did not account for all the densities. Seeding the GMM using the clustering method can lead to under-representation of the target molecule when there is significant non-target density in the reconstruction, e.g., lipid density in a membrane protein. This can be solved by either masking the density out before seeding the GMM, or increasing the number of Gaussian in the model.

In all examples shown in the paper, to avoid model bias, the GMMs were initialized directly from the voxel map, without the information from the existing PDB model. After initializing the coordinates of Gaussian functions, we initialize the amplitude and width for each Gaussian as constant values. The GMM is then optimized to match the voxel density map, using the same DNN-based procedure described in our previous paper[12]. All parameters, including coordinates, amplitude, and width are optimized at this step to match the GMM to the consensus density map. The optimization takes 40 epochs by default with a learning rate $10^{-6}$. After the optimization step, the output GMM typically has an average FSC greater than 0.9 at the target resolution range (Fig. S5).

**Global particle orientation refinement**

For the GMM-based global particle orientation refinement, we start from an existing homogeneous refinement and use the pose estimations provided by voxel-based methods. The protocol refines the orientation of each particle by aligning it to the GMM reference of the corresponding half set, and generates voxel-based density maps by reconstructing the particles at their new orientation. Conceptually, the GMM-based global orientation refinement protocol is similar to any classical voxel-based particle refinement protocol, in which particle images are aligned to reference structures iteratively. The main difference here is that GMMs, instead of voxel-based density maps, are used as alignment references.

If the initial particle orientation assignment is performed using the voxel-based EMAN2 refinement[6], it can be directly used for the GMM-based refinement. Otherwise, the particles and orientations can also be imported from other software[5,13]. When importing the particles, we read the initial splitting of the even/odd subset from the previous refinement, and keep the two subsets of particles separate throughout the refinement process. In addition to the orientation assignment, when the alignment is imported from other packages, we also read the existing CTF information from the star files, and flip the phase of particles according to the CTF during the import.

After all particles are imported, we reconstruct one voxel-based density map for each half set of particles, and build one GMM to match each of the half maps. The GMMs are then used as references for the local orientation optimization of the corresponding half set of particles. To refine the orientation of particles, we use a gradient based optimization algorithm[14] to find the best Euler angles and translation for each particle, so that the FRC between the particle image and the projection of the GMM is maximized. The optimization takes 10 iterations per particle, with a learning rate of $10^{-3}$. Since the particles are phase-flipped during import, and the FRC is normalized by the Fourier ring, we do not consider any effect of CTF in the comparison. Because the GMM is a continuous function in 3D, the orientation optimization can be performed using a gradient-based optimizer, where the gradient is computed by the automatic differentiation in TensorFlow.

After orientation refinement, one 3D density map is reconstructed from each half set of particles using the newly determined Euler angles. The "gold-standard" FSC curves are computed from the voxel-based density maps at the end of each iteration, and the maps are low-pass filtered according to the measured resolution. Finally, a new GMM is built from each half map, which is used as reference for the next iteration of refinement.

**Focused refinement**

The process of focused refinement, as well as the patch-by-patch refinement described below, is similar to global orientation refinement, with the addition of applying a mask to the GMM before the alignment process. This mask helps to focus the refinement on specific regions of interest.

The mask for focused refinement can be defined by the user to concentrate on particular areas, or can be generated automatically in a patch-by-patch refinement procedure. In each iteration of focused refinement, the mask is applied to the width of the Gaussian in the GMM. For a mask with values between 0 and 1, the width ($\sigma_j'$) of a Gaussian function centered at ($x$, $y$, $z$) is calculated using the expression:

$$\sigma_j' = \sigma_j/(0.25 + 0.75\, M_{x,y,z})$$

Here, $\sigma_j$ is the original width of the Gaussian, and $M_{x,y,z}$ represents the value of the given mask at the corresponding point. It's important to note that the width of the Gaussian is increased outside the mask, as larger width corresponds to lower local resolution, not necessarily lower density. This contrasts with an amplitude mask, which decreases the Gaussian amplitude outside the mask. The masking operation is applied only to Gaussian functions whose centers fall outside the mask. This ensures that the resulting GMM still has soft boundaries, even if the given mask has sharp edges. The width mask is applied to the GMM after it has been trained to match the reconstructed 3D density map at each iteration. This focused refinement process allows for the refinement of orientations in specific regions of interest, providing higher-resolution details in those areas.

**Patch-by-patch refinement**

In patch-by-patch refinement, we segment the full GMM into multiple patches, and perform focus refinement on each individual patch by increasing the Gaussian width of all the other patches. The results of focused refinement are then merged together to form a final composite map, which will have better resolved features throughout the entire structure.

To perform patch-by-patch alignment, we first divide the GMM into multiple patches using the K-means clustering algorithm. The number of patches (K) is determined empirically, and is set to 8 by default. In theory, the selection of the number should be based on the dynamics within the system, which is unknown before actually performing the refinement. If the number is set to 1, the process will become the same as global particle pose refinement. Increasing the number of patches leads to smaller target regions in each focused refinement, which can often result in improved resolution. However, the higher bound of the patch number is constrained by the computational power and the size of the protein. The focused refinement is on average as fast as the global orientation refinement, but it takes K times longer to run the patch-by-patch refinement with K patches. Additionally, if patches are too small, there may not be enough features within a patch for the alignment. In practice, to avoid aligning to noise, we always keep the size of a patch larger than a single α-helix or β-sheet. In our tests, the default number of K always yields an improvement of resolution and resolvability local features. To determine the best number of patches for a specific dataset, it is recommended to run multiple tests and decide empirically the number based on the results.

After segmenting the GMM into clusters, we create a soft spherical mask for each cluster of Gaussian functions that covers the center of all Gaussian coordinates in the cluster. Each mask is centered at the center of its corresponding GMM cluster, and the radius of the sphere is set to be the distance of the farthest Gaussian coordinate in that cluster from the cluster center. A five pixel soft falloff is added to each spherical mask, and adjacent masks have overlapping boundaries (**Fig.S7**).

For each patch, three iterations of focused refinement are performed using the corresponding spherical mask. After all focused refinement finishes, we combine the final voxel map reconstructions from each patch to form a composite, voxel-base, density map. This is done through weighted averaging using the patch masks:

$$Map_{composite} = \left(\sum_{i=1}^{K} Mask_i \cdot Map_i\right) / \left(\sum_{i=1}^{K} Mask_i\right)$$

Here $K$ is the number of patches, $Mask_i$ is the spherical mask of each patch, and $Map_i$ is the unmasked and unfiltered voxel-based reconstruction from the focused refinement of that patch. One $Map_{composite}$ is generated from each half set of particles, and a final "gold-standard" FSC curve is calculated from the two composite half maps.

**Conformation to pose conversion**

The conformation of particles is estimated by the GMM and DNN-based heterogeneity analysis[12,15]. After the analysis, the trained encoder network can output one set of coordinates for each particle image in the latent space, and the decoder network generates the GMM at the corresponding conformation based on the latent space coordinates. To convert the conformation of each particle to a change of pose, we first generate one GMM for each particle using the trained decoder. Then, to estimate the new particle pose under a given mask of the target region, for each particle, we search for a new angle that minimizes:

$$RMSD(\ (Project(GMM_{neutral}, \theta_{init}) - Project(GMM_{conf}, \theta_{new})) \cdot mask)$$

Here, $Project(GMM, \theta)$ makes the projection of a 3D GMM to 2D at the Euler angle $\theta$. $GMM_{neutral}$ is the GMM built from the consensus refinement, and $GMM_{conf}$ is the GMM at the conformation of that particle. $\theta_{init}$ is the initial pose assignment of the particle, and $\theta_{new}$ is the new pose to be optimized. The region of focus is specified by $mask$, which is a vector with length of the number of Gaussian functions in the GMM. The $mask_i$ is 1 if the $i$th Gaussian falls within the target region, and 0 if outside the region. Root Mean Square Deviation (RMSD) is used to measure the difference between the two GMMs. The optimization is done using the Adam optimizer through TensorFlow[14], starting from the initially assigned angle for each particle ($\theta_{new} = \theta_{init}$). For each particle, the optimization takes 30 iterations, and a learning rate of $10^{-3}$ is used.

As described in our previous work[12], at the beginning of the heterogeneity analysis, the decoder is pre-trained so that an input of zero vector will lead to an output of the GMM fitted to the consensus reconstruction. Therefore, the origin of the latent space from the heterogeneity analysis always corresponds to the conformation of the neutral GMM. By aligning the target region of the GMM at any conformational state to the GMM at the origin, we essentially transform the frame of reference to the target domain. That is, we switch from a system with a rigid core and a moving domain, to a frame of reference that the target domain is staying still, whereas the rest parts of the protein are moving. By reconstructing the particles at the new orientation, we can obtain a structure that is well resolved at the target domain and smeared out everywhere else.

After reconstructing the particles at the new orientation, we perform multiple iterations of focused refinement with the same mask. To avoid overfitting, the heterogeneity analysis is done for the even/odd particle subsets independently, with one pair of encoder-decoders built for each subset. While the latent spaces produced by the two encoders do not necessarily match, as long as the two neutral state GMMs are aligned to each other, the reconstructions of particles at the converted pose will still be aligned. Since we finalize the result using focused refinement, the structure from the conformation to orientation conversion can be simply treated as one single patch in the patch-by-patch refinement, so it can be merged with the results from all other patches and produce the final composite map with better features globally.

**Maintaining "gold-standard" validation**

To avoid overfitting, the entire process of GMM-based refinement and heterogeneity analysis is carefully designed to follow the "gold-standard" validation. From the beginning of the workflow, the two subsets of particles, as well as the corresponding reconstructions and GMMs are kept separate from each other in every step of the protocol.

When importing the initial alignment from other software packages, we preserve the even/odd split of particles from the upstream analysis. The initial half maps are generated by reconstructing particles of the corresponding half set at their given orientation. Two consensus GMMs are built from the two half maps independently, in order to make sure no bias from original reconstruction is introduced.

For the global orientation and focused refinement, the same "gold-standard" protocol is followed as any voxel-based refinement routine. Two GMMs are always built independently from the half maps, and they are only used as reference for the orientation search of particles from the corresponding half set. After alignment, 3D voxel maps are reconstructed from particles at the new orientation, and FSC curves are calculated from those voxel reconstructions. All FSC curves, except for the map-model FSC in **Fig.S5**, are "gold-standard" FSC curves between the two voxel-based half maps reconstructed from the particles.

In patch-by-patch alignment, for each patch, we use the same soft, spherical mask for both subsets of particles, so that the structures from individual focused refinement of the two subsets are comparable, and the "gold-standard" FSC can be computed. During each focused refinement, each particle is only aligned to its corresponding half map, and the resolution is measured by comparing the focused refinement results of the two half sets of particles using the same patch mask. After focused refinement of all patches finishes, we generate one composite map from each half set of particles, and the final FSC is computed from the two composite maps.

For the heterogeneity analysis, to make sure we do not introduce model bias and over-estimate resolution, we build one pair of encoder-decoder DNNs completely independently for each half set of particles. Here, to be absolutely sure we are not introducing model bias, we follow a stricter version of the "gold-standard" validation than many of the typical CryoEM data processing procedures. In most CryoEM data analysis, the heterogeneity of the entire dataset is analyzed using classification[21] or manifold method[22] before splitting the dataset to two halves and performing the "gold-standard" refinement. By doing heterogeneity analysis independently on two half sets and still reaching higher resolution, we show that the two subsets of particles not only agree on the consensus structure, but also share the same pattern of dynamics.

**Structure comparison**

For each dataset, we start from importing the extracted particle sets from EMPIAR provided by the authors. First, CTF correction is done by phase flipping each particle according to the information in the corresponding metadata (Relion star file)[5]. The Euler angles, 2D translation, and the "gold-standard" subset assignment of the particles are also extracted from the metadata, to use as the starting point of the GMM-based refinement. 3D reconstruction is performed using direct Fourier inversion from the particles at their assigned orientations. To ensure the conversion is correct, a script is also provided that can convert the Euler angle GMM-based refinement back to the star file. The metadata can then be converted to the format of CryoSPARC or other software using the Pyem package[23]. Reconstructions in the other software packages using the GMM-refined Euler angles also show improved FSC curves and real space features.

Compared to voxel-based refinement results, the global GMM-based orientation refinement changes the Euler angle assignment of the particles by 2-3 degrees on average, and the translation change is often less than one pixel. From the GMM-based global orientation refinement to focused refinement of individual domains, the orientation assignment change would depend on the flexibility of that domain. For example, in the focused refinement of the transmembrane part of the GPCR, the Euler angle assignment of particles changes 3.7 degrees on average (**Fig.S6**).

One soft mask is created for each dataset and used for the "gold-standard" FSC calculation. To avoid masking induced artifacts when comparing the FSC measurement, masks used here are often not as tight as the ones used in the original publications, leading to slightly lower resolution numbers than reported. However, real space features are essentially identical since the maps are filtered by their gold-standard resolution locally.

To ensure the real space features between different maps are comparable, we use the same sharpening method for all datasets. For each structure, we simulate a density map from the corresponding molecular model provided by the authors at 2Å, then compute the radial Fourier amplitude profile of the simulated map. The amplitude profile is used to sharpen the reconstructions before they are low-pass filtered to the determined resolution.

The final density maps are low-pass filtered to their gold-standard resolution locally using the method implemented in EMAN2[24]. The corresponding molecular model from the original publication is fitted to the density map, and automated real space refinement protocol from Phenix is used to optimize the model to fit into the refined map locally[25]. Finally, Q-score is computed using the real space refined model and the map, and the averaged Q-score per residue, smoothed over a 11 residue window[18], is reported in the figures. For the local feature comparison, we choose regions with the largest difference between the Q-score from the initial and refined reconstructions. Display of 3D volumes is done in UCSF Chimera and ChimeraX [26,27].

**Detail on datasets**

For the GPCR dataset, we use picked particles of the P-Neurokinin Receptor G protein complexes from EMPIAR-10786[17]. Specifically, the reconstruction and final Euler angle assignment of the SP-NK1R-miniGs399 is used. The dataset includes 288,659 particles with a pixel size of 0.86Å. According to the original publication, the orientation assignment uses a combination of Relion, CryoSPARC and CisTEM. It is worth noting that the original structure is also the result of a final round of focused refinement, which targets the substrate binding region. After the GMM based refinement, we improve features on both the region targeted by the focused refinement, as well as domains at the other end of the protein. The GMM is seeded on the protein density only, excluding the lipid shell, and the FSC is evaluated using a larger soft mask that includes both the protein and lipid density. 7,374 Gaussian functions are included in the GMM, which is the number of non-H atoms in the corresponding molecular model. The "gold-standard" resolution of the initial structure was 3.3Å, which was improved to 2.8Å after the GMM-based global refinement, and 2.5Å after patch-by-patch refinement. The molecular model (PDB: 7RMH) corresponding to the EMPIAR entry was fitted into the density map using Phenix real space refinement against the final density map from the patch-by-patch refinement.

For the dataset of SARS-COV2, we use picked particles from EMPIAR-10492[19], which includes 29,180 particles with their corresponding metadata. The original Euler angle assignment was performed using Relion. Here, since the resolution is lower, we use a GMM of 17,000 Gaussian functions. The number of Gaussian is estimated so that the GMM matches the density map at the target resolution range. The pixel size of the particles is 1.061Å, and the structure is determined with c3 symmetry. The symmetry is expanded to c1 for the focused refinement by duplicating each particle three times at the three symmetrical orientations, and the mask for the refinement covers the target domain of one asymmetric unit. To generate the final composite map, we reimpose the c3 symmetry to the focused refinement results by weighted average, using the focus masks as weights, then merge the maps from different focused refinement together using the same technique. The "gold-standard" resolution of the initial structure was 3.8Å, which was improved to 3.4Å after the GMM-based global refinement, and 3.1Å after patch-by-patch refinement. The molecular model (PDB: 6ZWV) was fitted into the density map using Phenix real space refinement against the result of patch-by-patch refinement.

The ABC transporter dataset comes from EMPIAR-10374[20], a human ABCG2 transporter with inhibitor MZ29 and 5D3-Fab. The dataset contains 284,831 particles with a pixel size of 0.84Å, and the initial orientation is determined using a combination of CryoSPARC and Relion. The GMM includes 15,174 Gaussian functions, the same as the number of non-H atoms in the corresponding molecular model. The Euler angles from Refine3d_C2 are used as the initial particle orientation assignment, and the refinement is performed with c2 symmetry. Similar to the SARS-COV2 dataset, the symmetry is expanded for heterogeneity analysis and focused refinement, then reimposed to generate the final composite map. The GMM-DNN based heterogeneity analysis is performed focusing on the Fab, using a hierarchical GMM[15] and target 5Å resolution. The molecular models of different conformations (shown in **Fig.2B**) are generated by morphing the neutral state molecular model (PDB: 6ETI) using the decoder trained from the particles. Since the C-domain of the Fab is not modeled in the original publication, we use an existing Fab structure (PDB: 7FAB), rigid-body fitted to the density map, then optimized to match the final composite map using Phenix real space refinement. The length of the movement trajectory is measured at the residue with the longest movement distance, specifically 205K of the heavy chain. The average Q-score improves from 0.62 to 0.63 after the GMM-based global refinement, and 0.65 after we merge the patches together. Note that the resolvability of the Fab C-domain is not computed resulting in the absence of the Q-score in **Fig.S4**, due to the difficulty of fitting the molecular model into the original density map reliably to make a fair comparison.

**Performance of GMM-based refinement at medium resolution**

In the three examples above, we have already shown that the GMM-based refinement can improve the global "gold-standard" resolution of the entire protein when the initial structure was solved at 3-4Å resolution. In the example of the ABC transporter, we also showed that the GMM-based heterogeneity analysis and refinement can improve the local resolution of the Fab domain, which was initially at 5Å or lower resolution due to its high flexibility. In this section, we test the GMM-based refinement on datasets with lower global resolution, to answer the question whether the GMM-based method still improves the result when there is no high-resolution feature such as side chain densities in the initial reconstruction. To demonstrate the performance of the GMM-based refinement at lower resolution, we create small subsets of the GPCR and SARS-COV2 datasets, and compare the results of voxel and GMM-based refinement (**Fig.S8**).

In the GPCR dataset, we randomly selected 6,000 particles from the full dataset (3,000 from each half set) for the refinement. Using the small subset of particles, we performed "gold-standard" refinement from scratch using the voxel-based single particle refinement protocol in EMAN2[28]. The high resolution reconstruction was phase randomized to 10Å and used as the reference for the refinement. The final structure reached the global resolution of 4.5Å with 6,000 particles, according to the "gold-standard" FSC curve. After the voxel-based refinement, we continued to perform the GMM-based global and patch-by-patch refinement on the same dataset. The GMMs were built directly from the half maps from the voxel-based refinement, using the same protocol described for the processing of the full dataset. The number of Gaussians is estimated to be 7,000 from the density map. The GMM-based global refinement improved the "gold-standard" resolution to 4.0Å, and the patch-by-patch refinement further improved the resolution to 3.9Å with 6,000 particles. From the reconstructed maps, the GMM-based refinement also shows clear improvement of real space structural features. At the transmembrane domain, side chain densities start to show up from the previously smooth α-helices, and the strains in the β-sheets becomes separable at the lower part of the protein.

The processing of the small subset from the SARS-COV2 spike protein dataset follows the same protocol. 6000 particles were randomly selected from the full dataset for the refinement. The initial voxel-based refinement reached 6.2Å according to the "gold-standard" FSC curve. Similarly, the GMMs were built from the half maps, and the number of Gaussian is set to 12,000, which is directly estimated from the density map. The GMM-based global refinement improved the "gold-standard" resolution to 5.3Å, and the patch-by-patch refinement further improved the resolution to 4.6Å. Improvement of real space features can also be seen in the reconstructions after the GMM-based refinement, as the helical pitch starts to show up at the bundle of α-helices at the center of the spike protein.

Clearly, in both datasets, the resolution is limited by the number of particles in the subsets we created. By improving the resolution of the refinement from the small subsets, we show that the GMM-based refinement is applicable to a wide resolution range, and does not require the presence of high-resolution features, e.g., side chains, in the initial reconstructions.

**Computational resource consumption**

The alignment protocol is implemented using Tensorflow[16], and runs on GPUs with CUDA capability. In our tests, for the refinement of 400,000 particles with box size of 196 pixels and targeting 3Å resolution, one iteration of refinement takes ~2 hours on a Nvidia RTX A5000 GPU. For all examples shown in the paper, five iterations of refinement are sufficient to reach convergence by the FSC measurement. I.e., the "gold-standard" FSC curves do not show clear improvement from the 4th to the 5th iteration. Focused refinement has virtually the same time and resource cost as the refinement of the full protein, and the time consumption of patch-by-patch alignment is N times the focused refinement, where N is the number of patches.

During the refinement of the GPCR dataset, which uses a GMM of 7374 Gaussian functions and a box size of 240, the peak GPU memory use is 8.7GB. In contrast, the peak memory usage during the refinement of the ABC transporter, which uses a GMM with 15,174 Gaussian functions and particles with box size of 256 pixels, is 17.1GB. Note that the GPU memory usage is roughly linear to the batch size used during the refinement, i.e., the number of particles loaded into the GPU memory at the same time, which is set to 16 by default and can be adjusted by users. Different from the training of DNNs, here the pose of each particle is independent, and we do not need to consider the statistical power for each batch, so this number can be adjusted to balance the speed and GPU memory consumption. I.e., using a batch size of 8 will save about half of the GPU memory in exchange for longer runtime, but the refinement results will not be affected.

The CPU memory consumption during the refinement is also adjustable. By default, 20,000 particles are loaded into the CPU memory each time through the process, which can be changed through program options. However, unlike the batch size that affects the parallelism in GPUs, loading fewer particles into the CPU each time reduces the CPU memory cost but does not significantly slow down the process.

**Interpretation of FSC curves**

During the comparison of the structures, it is worth noting that the patch-by-patch refinement sometimes contributes less than expected to the global "gold-standard" FSC curves. For example, in the SARS-COV2 spike example (**Fig.S3**), the patch-by-patch refinement is performed on the top of the GMM-based global alignment, and improves the structure in virtually every aspect. This includes better local resolution throughout the protein, equal or higher Q-score for every residue, and improved real space features. However, the improvement of the overall FSC from the patch-by-patch refinement is relatively small, particularly when compared to the improvement from the original orientation assignment to the result from the global GMM-based refinement. To solve the puzzle, we use a simple simulated example to demonstrate the impact of the structural flexibility on the FSC curves.

Here, we generate simulated CryoEM density maps using the molecular model of β-galactosidase (PDB: 6CVM)[29]. To simulate the structural flexibility, we randomize the coordinates of each atom within a local region before turning the model into a density map. This is done by shifting each atom toward a random direction for a given distance, which essentially controls the local resolution around the atom. For each of the even/odd subset, we generate 10 maps with randomized coordinates, and average them together to produce the map for the half set. Gaussian noise is added to the half maps so that the average SNR is 0.5. The "gold-standard" FSC is then computed between the results of the two half sets, without masking or other post processing steps. For two out of the four subunits of β-galactosidase, we consider them the rigid "core" of the protein and introduce only small shifts, so the local resolution is ~3Å. For the other two subunits, we randomize the atom coordinates at different scales to simulate the impact of local structure flexibility. In the first example (blue in **Fig.S9**), we randomize the atom coordinates so the local resolution at the two flexible subunits is ~15Å, whereas in the second example (red), the local resolution at the same two subunits is ~7Å. Obviously, the two structures are essentially the same at the two rigid subunits, and the second structure is much better resolved than the first one at the two flexible subunits. However, the first structure shows a slightly better FSC curve, despite its worse resolvability and local resolution.

A simple explanation of the phenomenon is that the real space signal at different subunits is weighted differently in the overall FSC curves, since the FSC is computed in the Fourier space. In the first structure, while the flexible subunits are poorly resolved, the average voxel intensity of the two subunits are also much lower compared to the two rigid ones, since structures of different "conformations" are averaged together. As a result, these two subunits carry a very low weight in the FSC curve, and the structure can achieve a "gold-standard" resolution of 3Å despite having half of the map only resolved at 15Å. In the second example, while the resolvability of the flexible subunits is considerably improved compared to the first structure, it is still not as good as the rigid subunits. However, the two flexible subunits now have higher voxel intensity, due to the fact that they are better resolved, thus a higher weight in the overall FSC curve. This means, the high resolution parts of the density map now weigh less in the calculation of FSC, leading to a slightly lower curve than the first example.

In summary, this simulated example suggests that the resolvability improvement of local flexible domains does not necessarily lead to the improvement of the overall FSC. To have a better measurement of the quality of a CryoEM density map, it is necessary to combine metrics including the global FSC, local resolution, local resolvability measurement such as the Q-score, as well as the real space features.

**Software availability**

All computational tools described here are implemented in EMAN2, a free and open source software for CryoEM/CryoET imaging processing. The code is available at [github.com/cryoem/eman2](github.com/cryoem/eman2), and a tutorial for these tools can be found through eman2.org/e2gmm_refine.

**Data accessibility**

The three data sets used in the paper are publicly available through EMPIAR. EMPIAR-10786 for the GPCR, EMPIAR-10492 for the SARS-COV2 spike, and EMPIAR-10374 for the ABC transporter. Structures produced in this paper are deposited in EMDatabank, and the corresponding refined models in the PDB.

**Acknowledgements**

This research has been supported by NIH grant R21MH125285 and R01GM113195. Computational resources from SLAC Shared Scientific Data Facility (SDF) are used for the work.

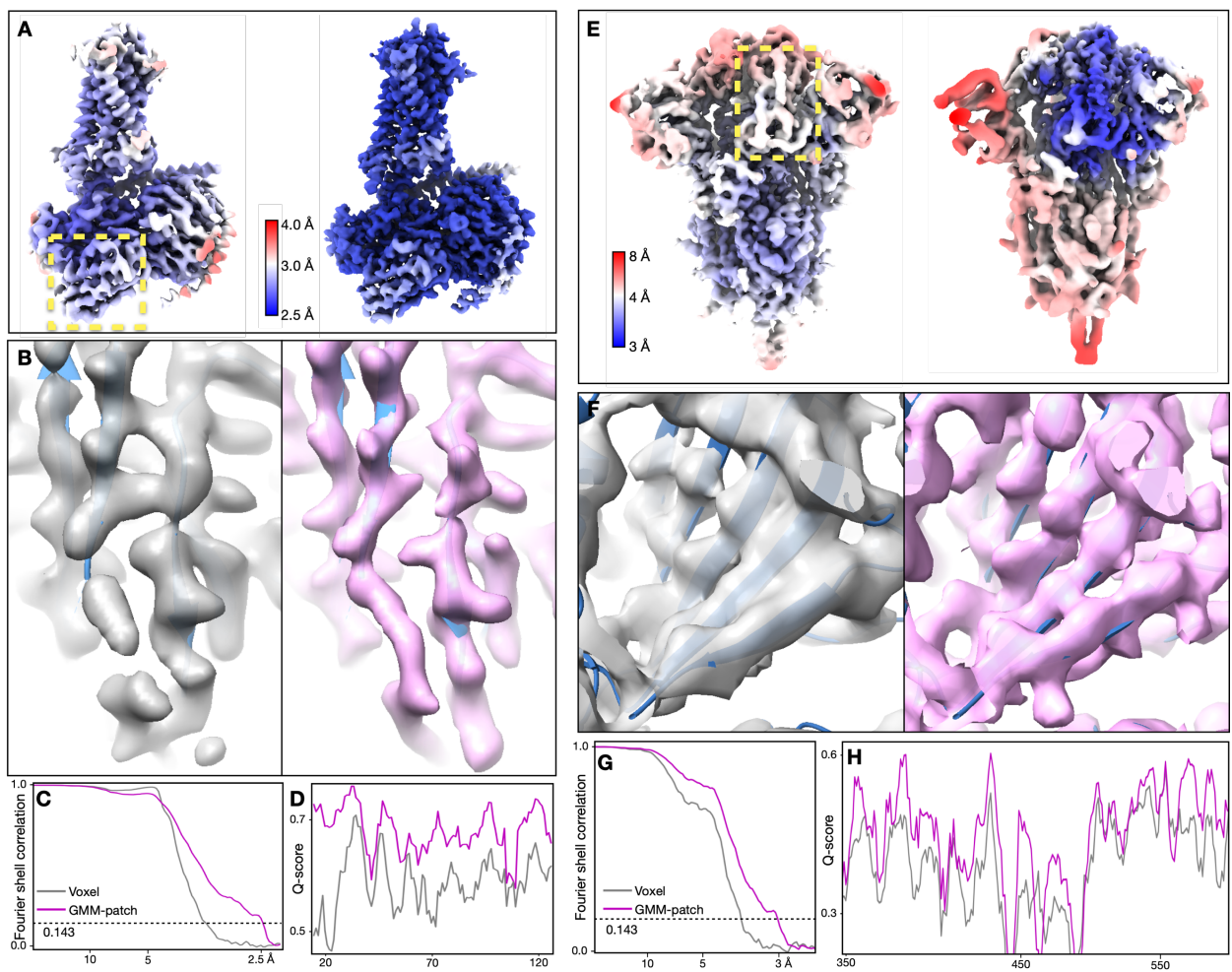

**Figure 1.** Refinement of GPCR (EMPIAR-10786) and SARS-COV2 spike (EMPIAR-10492) datasets. (A-B) Reconstructions of the GPCR using original angle assignment (left) and after the patch-by-patch alignment (right), colored by local resolution. The yellow box highlights the region shown of the corresponding real space features in B. (C-D) Comparison of real space features, "gold-standard" FSC and Q-score between the original reconstruction (gray) and the patch-by-patch refinement (pink). (E) Reconstructions of the SARS-COV2 spike using original angle assignment (left) and after the focused refinement of RBD (right), colored by local resolution. The yellow box highlights the target domain for focused refinement. (F-H) Comparison of real space features, "gold-standard" FSC and Q-score of RBD between the original reconstruction (gray) and the patch-by-patch refinement (pink).

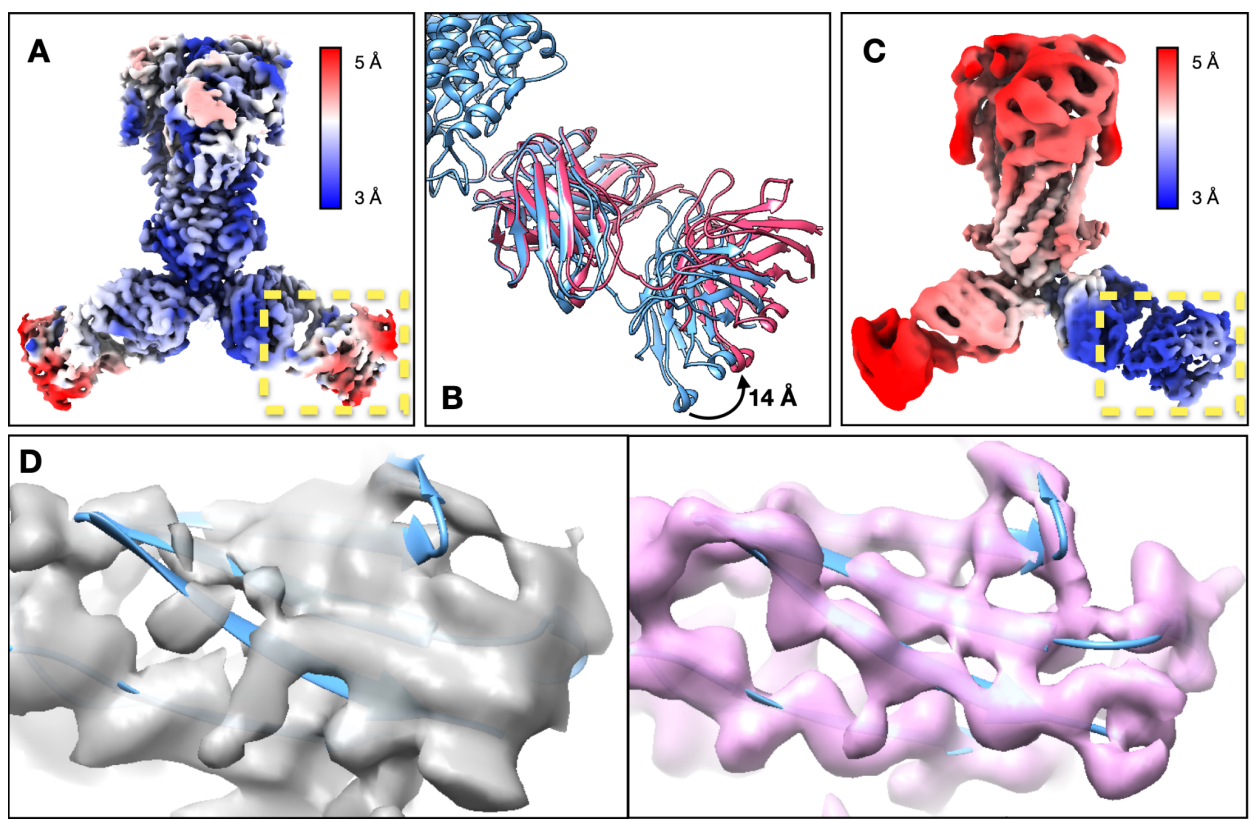

**Figure 2**. Refinement of the ABC transporter (EMPIAR-10374). (A) Reconstruction of the ABC transporter using original orientation assignment, colored by local resolution. (B) Continuous movement of the Fab shown in morphed molecular models. (C) Reconstruction of the ABC transporter after converting the movement trajectory to particle orientations, colored by local resolution. (D) Feature comparison at a β-sheet in the C-domain of Fab from the original structure (gray) and the structure after GMM-based refinement (pink).

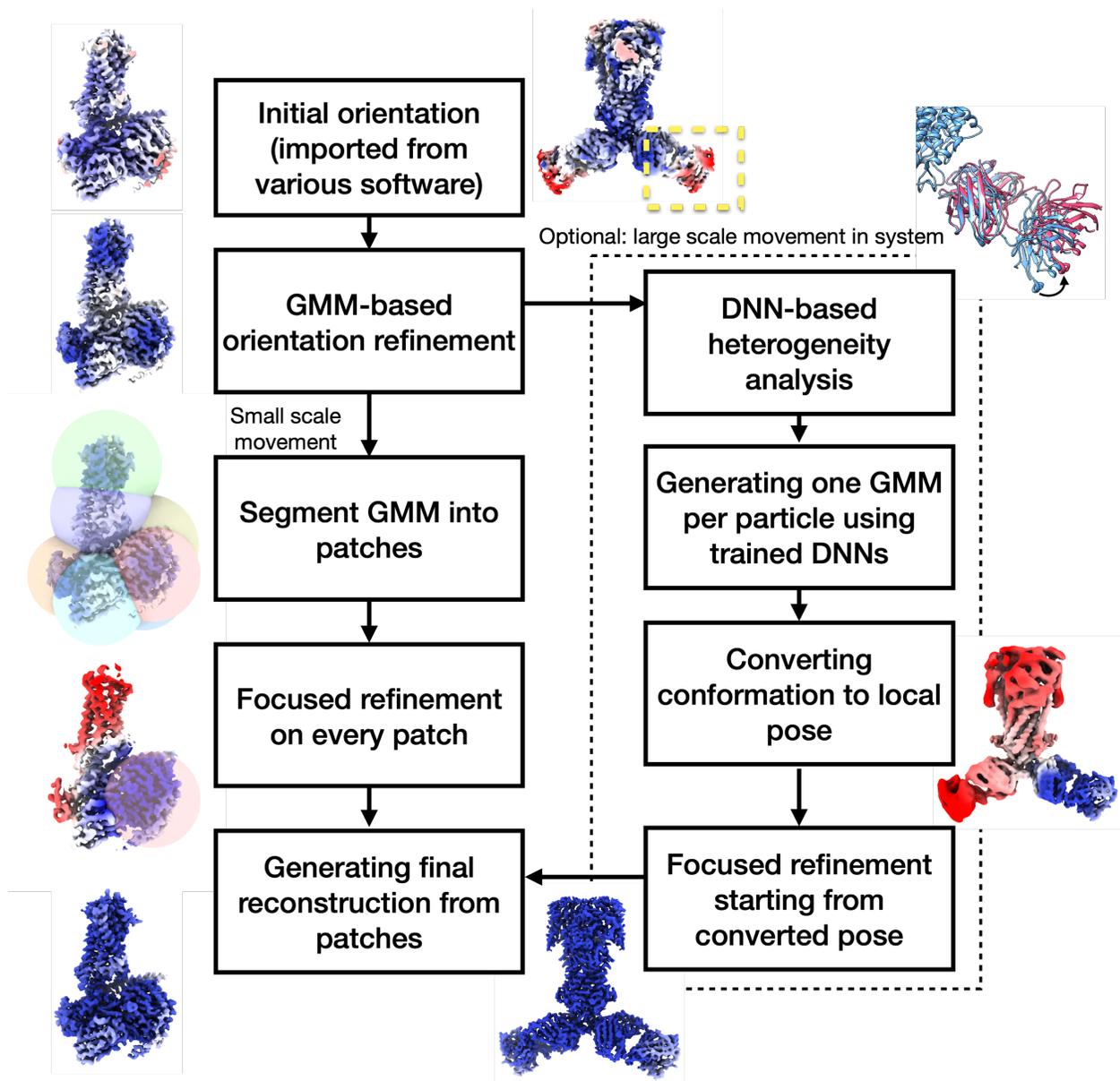

**Fig S1**. Workflow diagram for GMM-based particle orientation and conformation refinement. Each block represents one step of processing, and the arrows indicate the sequence of the processes. The right side, DNN-based heterogeneity analysis is optional and is only recommended when large scale movement is present in the system. The result of the DNN-based refinement focusing on one region can be treated as one patch and merged into the patch-by-patch refinement results from the main workflow.

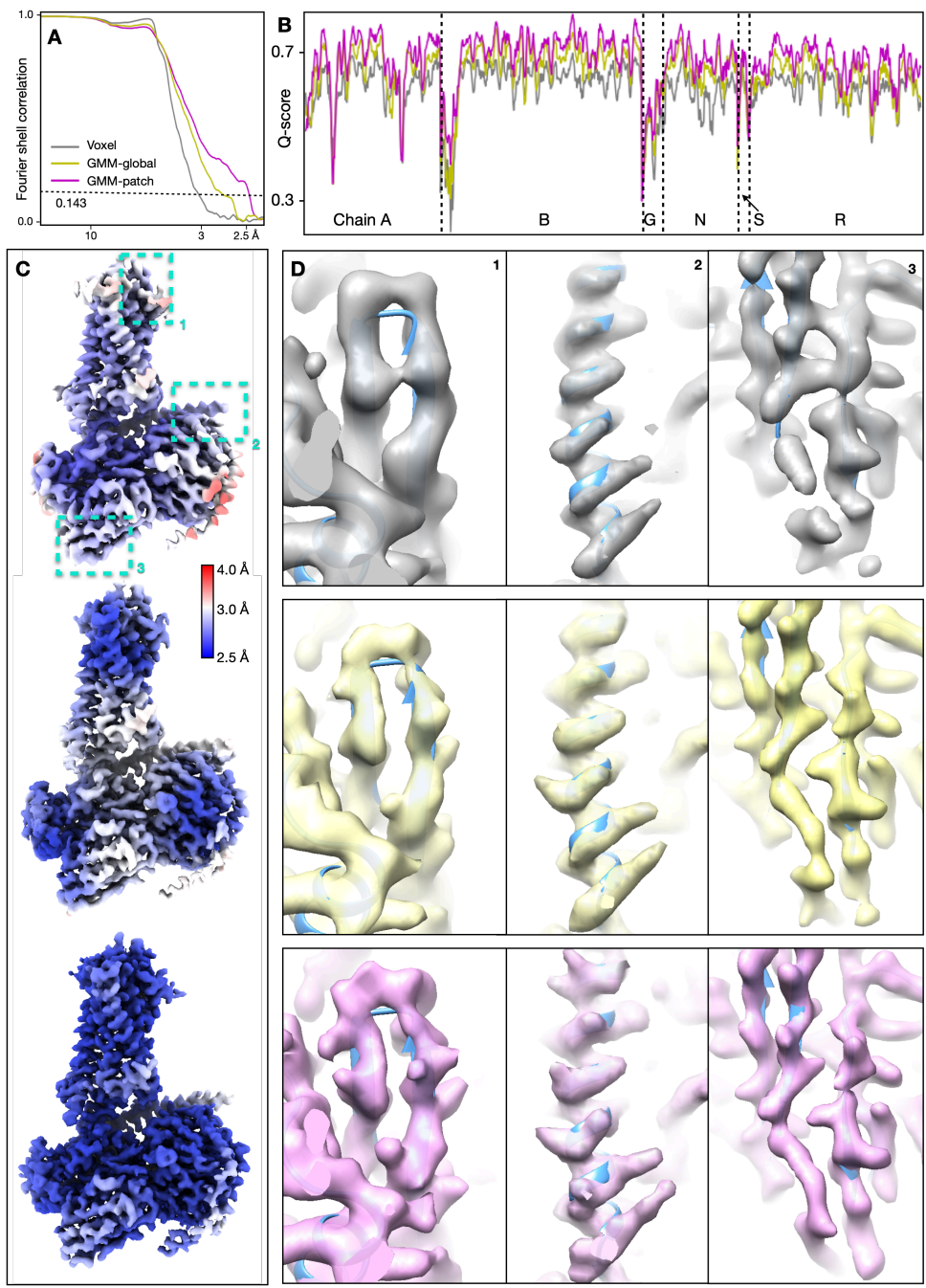

**Fig S2**. Detailed results of the GPCR dataset (EMPIAR-10786). (A) "Gold-standard" FSC curves of the reconstruction using initial orientation (gray), global GMM-based refinement (yellow) and patch-by-patch refinement (pink). (B) Q-score comparison of the three corresponding maps. (C-D) Overall structure of the three reconstructions as described in A and B, colored by local resolution, and comparison of local real space features. The cyan boxes in C highlight the location of features in the corresponding columns shown in D.

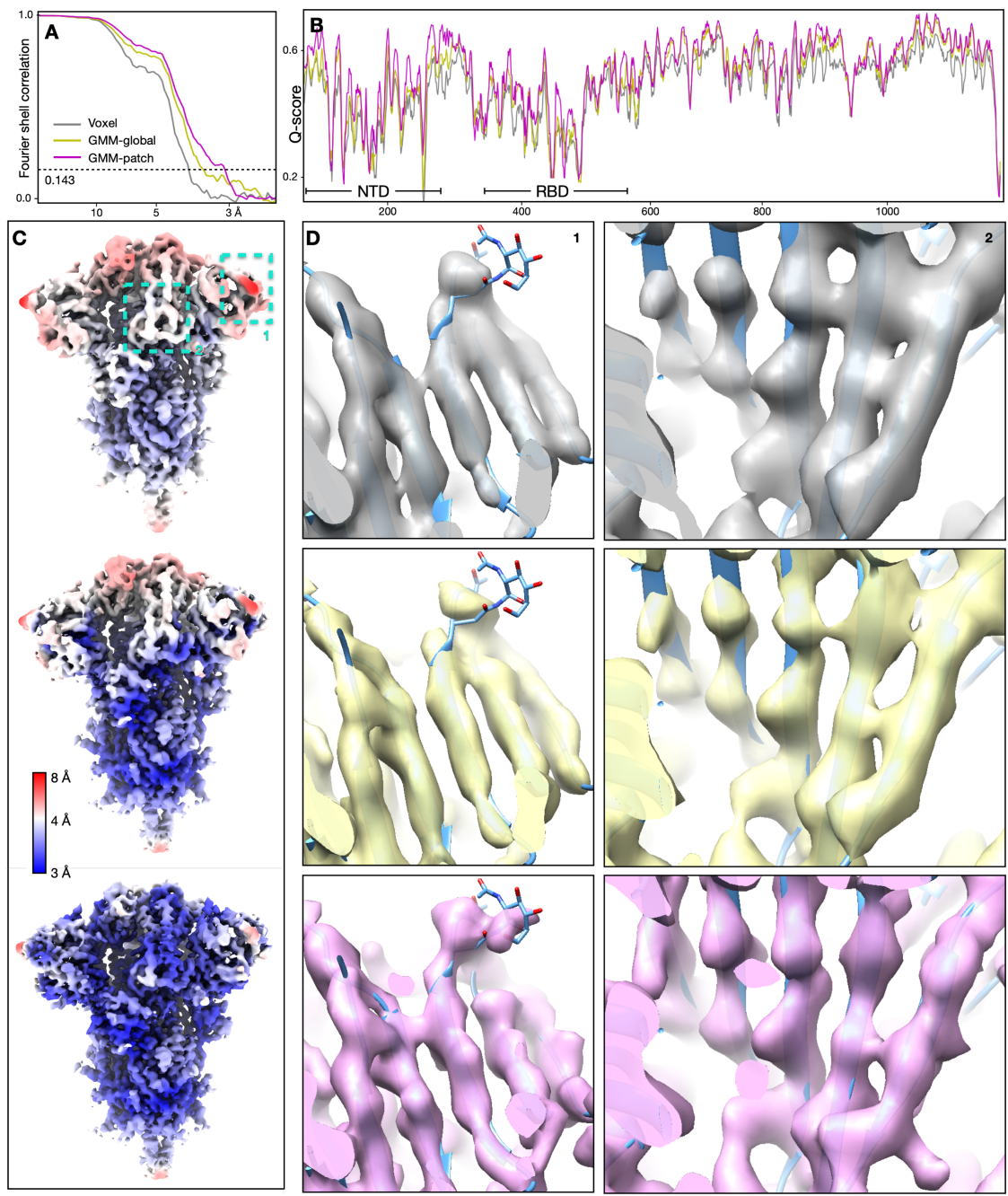

**Fig S3**. Detailed results of the SARS-COV2 dataset (EMPIAR-10492). (A) "Gold-standard" FSC curves of the reconstruction using initial orientation (gray), global GMM-based refinement (yellow) and patch-by-patch refinement (pink). (B) Q-score comparison of the three corresponding maps. (C-D) Overall structure of the three reconstructions, colored by local resolution, and comparison of local real space features. The cyan boxes in C highlight the location of features in the corresponding columns shown in D.

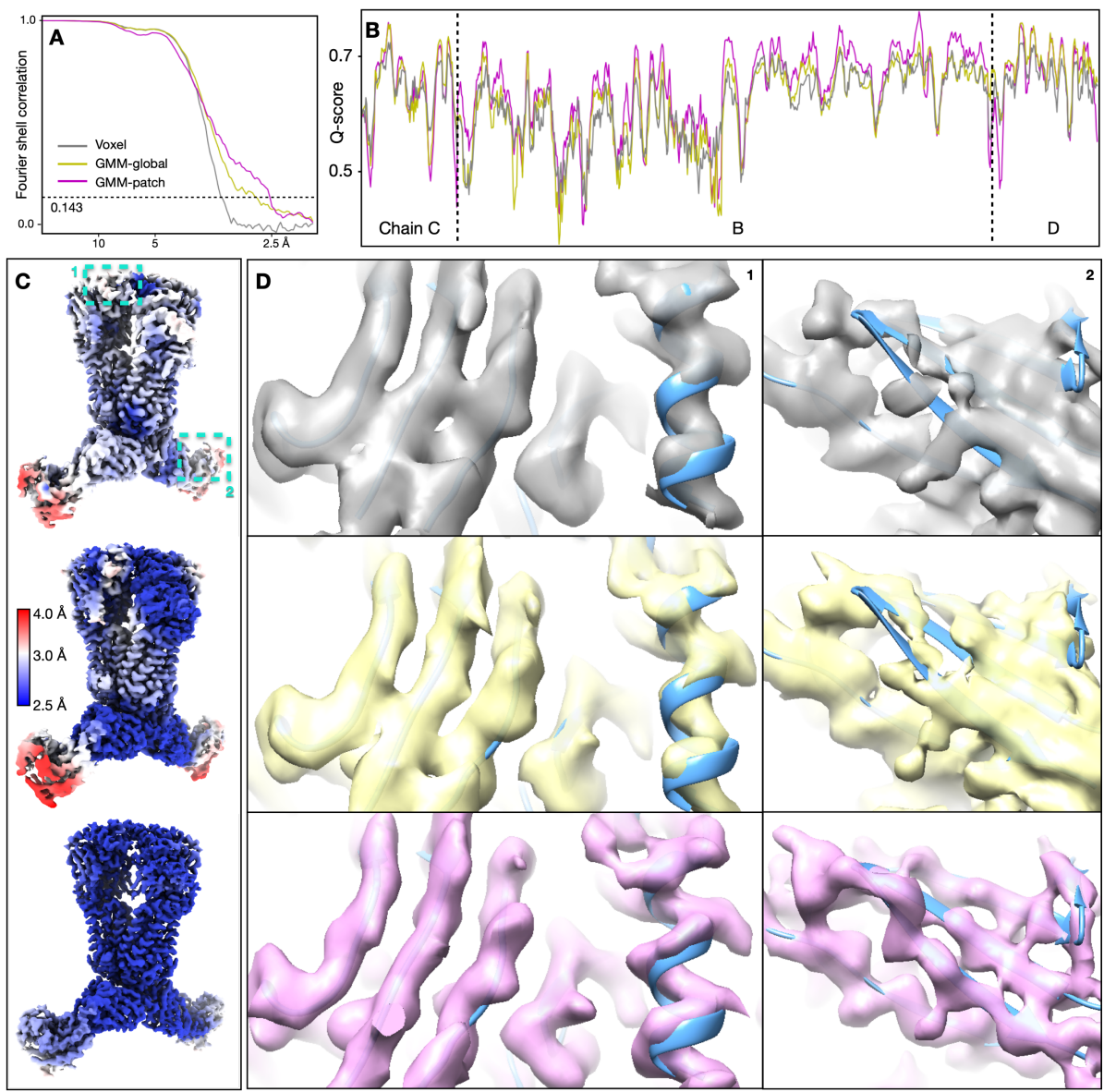

**Fig S4**. Detailed results of the ABC transporter dataset (EMPIAR-10374). (A) "Gold-standard" FSC curves of the reconstruction using initial orientation (gray), global GMM-based refinement (yellow) and patch-by-patch refinement (pink). (B) Q-score comparison of the three corresponding maps. (C-D) Overall structure of the three reconstructions, colored by local resolution, and comparison of local real space features. The cyan boxes in C highlight the location of features in the corresponding columns shown in D.

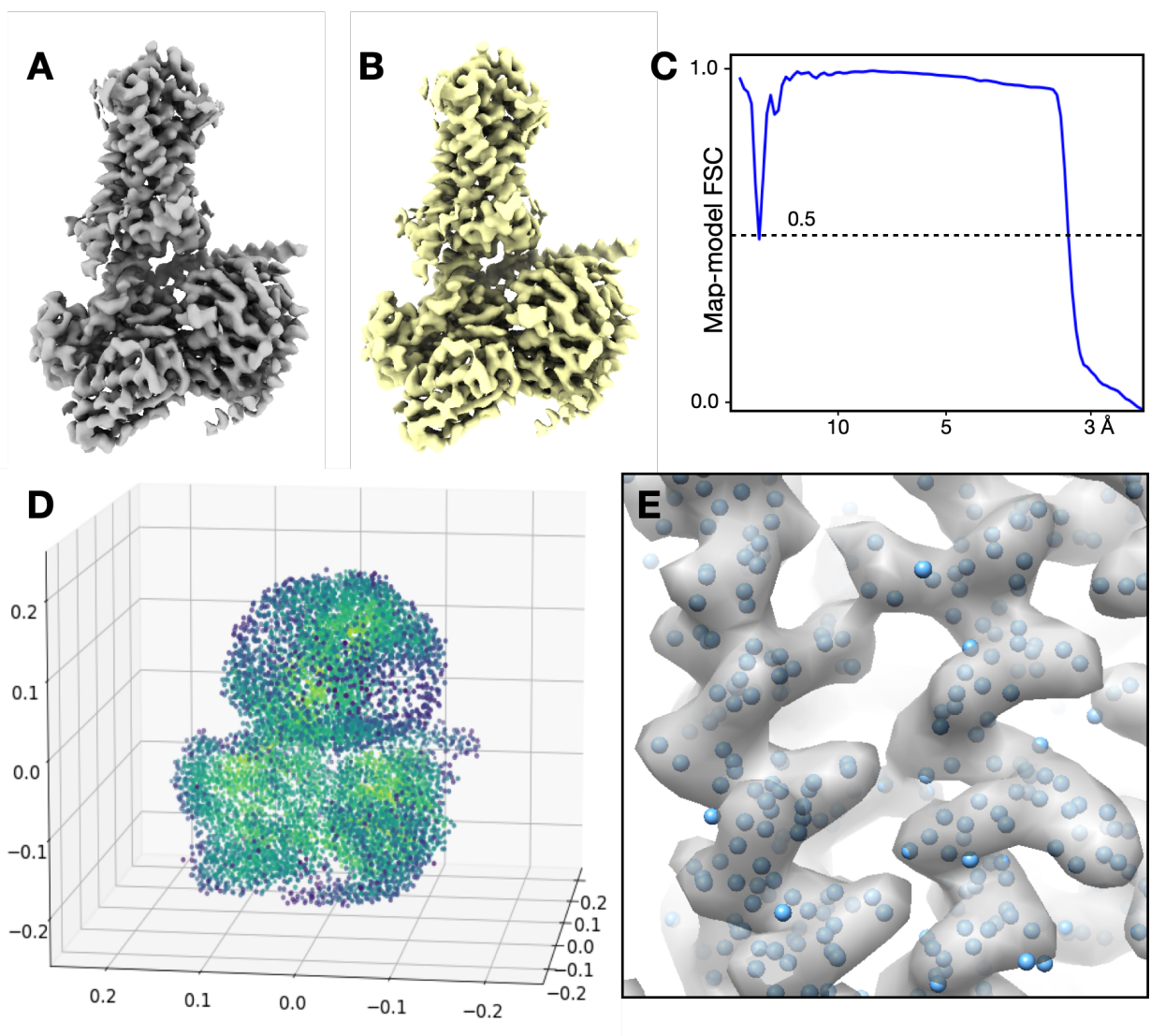

**Fig S5**. Visualization of GMMs. (A) Reconstruction of the GPCR using the voxel map representation, determined at 3.3Å resolution. $6 \times 10^6$ floating point values are required to represent the structure. (B) GMM representation of the A, shown in isosurface view. $5 \times 10^4$ floating point values are used to represent the structure. (C) FSC curve between A and B. The two structures are virtually identical up to 3.3Å. (D) Visualization of the GMM from B using 3D scatter plot. Each point is colored by the amplitude of the Gaussian function and the size of the points correspond to the width of Gaussian functions. (E) Overlay of the coordinates of Gaussian functions in the corresponding density map.

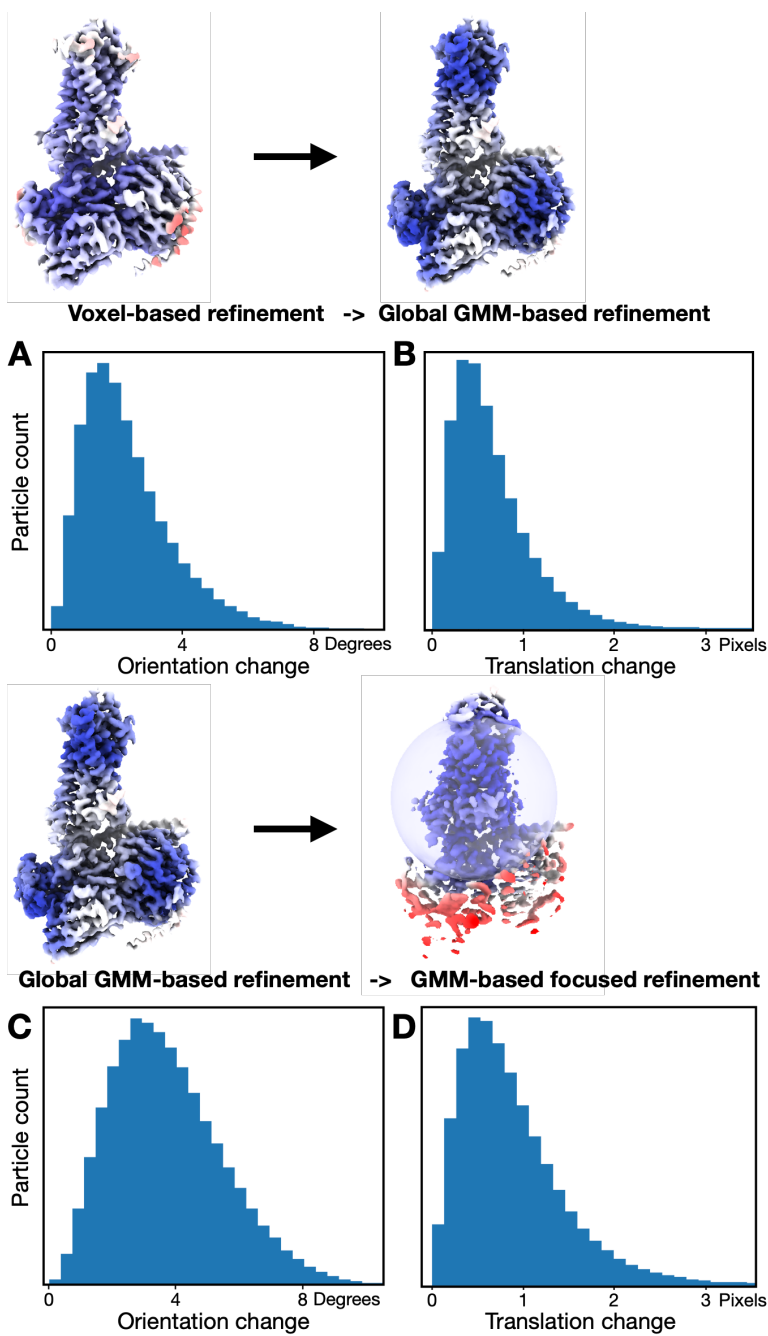

**Fig S6**. Orientation and translation change before and after GMM-based refinement. (A) Histogram of particle orientation assignment change after the GMM-based global refinement. Mean=2.34, std=1.37 degrees. (B) Histogram of particle translation change after the GMM-based global refinement. Mean=0.66, std=0.46 pixels. (C) Histogram of particle orientation assignment change after the GMM-based focused refinement. Mean=3.74, std=1.76 degrees. (D) Histogram of particle translation change after the GMM-based global refinement. Mean=0.89, std=0.62 pixels.

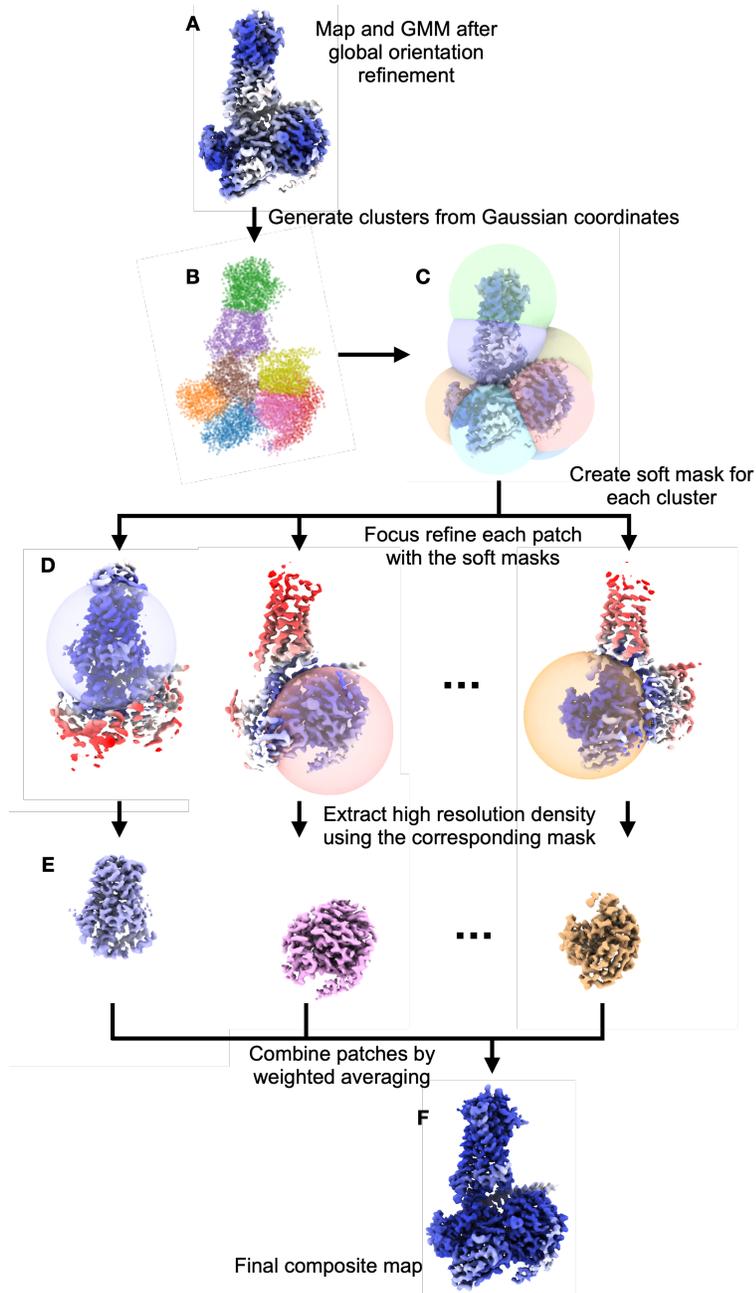

**Fig S7**. Diagram for the patch-by-patch refinement. (A) Input density maps and GMMs from the global orientation refinement. (B) Scatter plot of Gaussian coordinates, colored by clustering result. (C) Soft masks for each cluster/patch. Each mask is a sphere covering all Gaussian coordinates of the corresponding cluster, with a soft falloff. (D) Focused refinement results using the soft masks, colored by local resolution. Note that the same refinement process is done independently for the even/odd subsets of particles using the corresponding half map/GMM as reference. (E) Masked out density from the individual focused refinement result. (F) Final composite map generated by weighted averaging.

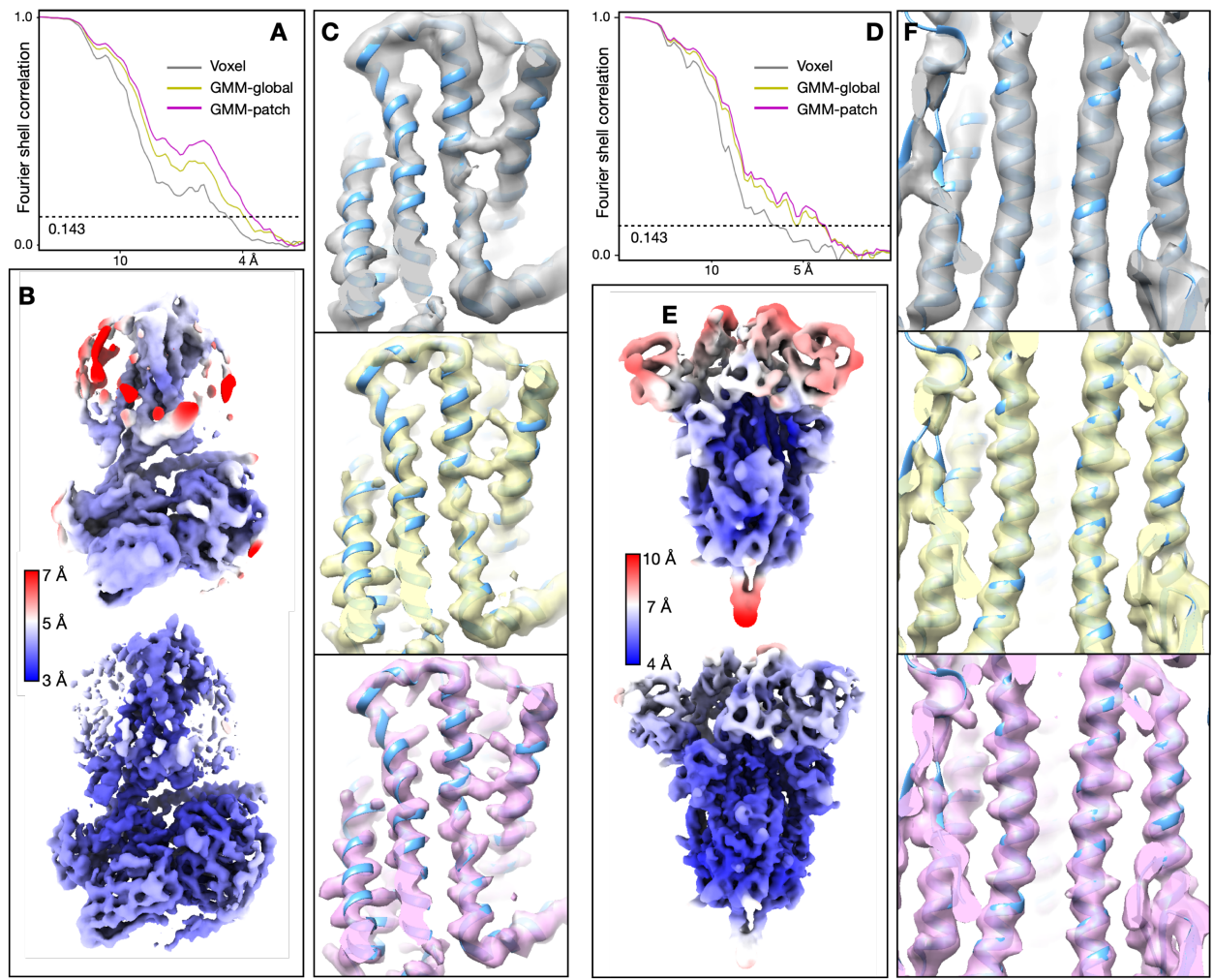

**Fig S8**. Performance of GMM-based refinement in datasets at lower resolution. (A-C) Refinement of a small subset of the GPCR dataset. (D-F) Refinement of a small subset of the SARS-COV2 dataset. (A, D) "Gold-standard" FSC curves of the reconstruction using voxel-based refinement (gray), global GMM-based refinement (yellow) and patch-by-patch refinement (pink). (B, E) Overall structure of the initial reconstruction and the final patch-by-patch refinement result, colored by local resolution. (C, F) Comparison of local features from the voxel-based, global GMM-based and patch-by-patch refinement.

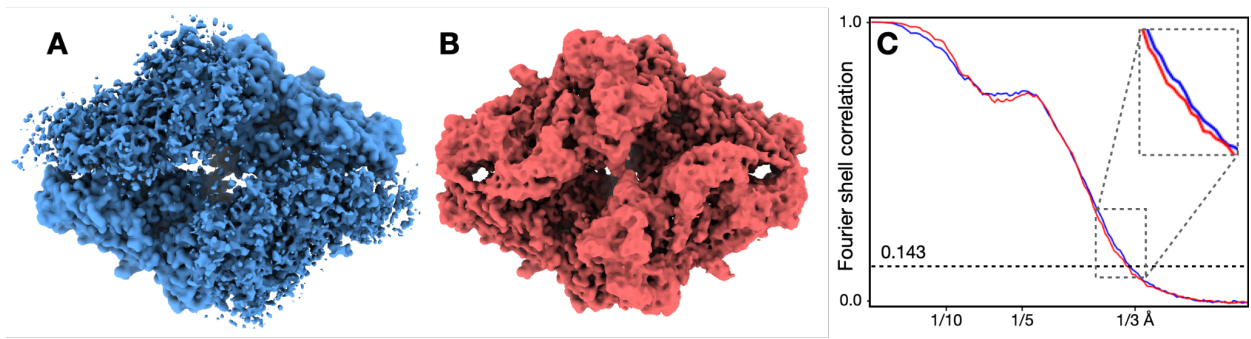

**Fig S9**. Impact of flexible domains on the global FSC curve. (A) Simulated density map of β-galactosidase, with two rigid subunits at 2.5Å and two flexible ones resolved at 15Å. (B) Same simulated map as A, with the two flexible subunits resolved at 7Å. (C) Comparison of "gold-standard" FSC curves of the two structures. Blue - A, red - B.

**Supplementary movie 1**

Structure comparison of the GPCR dataset (EMPIAR-10786). Gray: reconstruction using initial orientation. Pink: reconstruction after GMM-based patch-by-patch refinement.

**Supplementary movie 2**

Structure comparison of the SARS-COV2 dataset (EMPIAR-10492). Gray: reconstruction using initial orientation. Pink: reconstruction after GMM-based patch-by-patch refinement.

**Supplementary movie 3**

Structure comparison of the ABC transporter dataset (EMPIAR-10374). Gray: reconstruction using initial orientation. Pink: reconstruction after GMM-based patch-by-patch refinement.

**Supplementary movie 4**

Converting structural heterogeneity to particle orientation focusing on the Fab of the ABC transporter (EMPIAR-10374).